\documentclass[12pt,preprint]{aastex}

\shorttitle{Optical Variability of J~1128+5925}
\shortauthors{Wu et al.}

\received{}
\accepted{}

\begin{document}

\title{Optical Variability of the Radio Source J~1128+5925}

\author{Jianghua Wu, Xu Zhou, Jun Ma, Zhenyu Wu, Zhaoji Jiang, Jiansheng Chen}
   \affil{National Astronomical Observatories, Chinese Academy of Sciences,
        20A Datun Road, Beijing 100012, China}
   \email{jhwu@bao.ac.cn}

\begin{abstract}
Very recently, J~1128+5925 was found to show strong intraday variability at
radio wavelengths and may be a new source with annual modulation of the
timescale of its radio variability. Therefore, its radio variability can be
best explained via interstellar scintillation. Here we present the properties
of its optical variability for the first time after a monitoring program in
2007 May. Our observations indicate that in this period J~1128+5925 only
showed trivial optical variability on internight timescale, and did not show
any clear intranight variability. This behavior is quite different from its
strong radio intraday variability. Either this object was in a quiescent
state in optical in this period, or it is intrinsically not so active in
optical as it is in radio regimes.
\end{abstract}


\section{INTRODUCTION}
Blazars are the most variable subset of AGN. They show a variety of
variability timescales. The longest timescales can be far longer than one
year, while the shortest may be less than one hour.
The variability with a timescale less than one day is often called
intraday variability or IDV, as first reported by \citet{heeschen84},
\citet{witzel86}, and \citet{heeschen87}. Strong IDV phenomena have been
observed in the radio domain in a large number of blazars. If interpreted
as being source intrinsic, the short-timescale variability would require a
very small emitting region and hence a very high apparent brightness
temperature of $10^{16}\sim10^{21}\,\rm{K}$, which is far beyond the
inverse-Compton limit of about $10^{12}\,\rm{K}$ \citep{keller69}. 

Alternatively, the IDV can be explained via extrinsic origin, e.g., via
interstellar scintillation (ISS). A strong support to the ISS origin
is the so-called annual modulation of the variability timescale, which
is the result of the annual changes of the relative velocity vector between
the scattering screen and the Earth as the Earth orbits around the Sun
\citep[e.g.,][]{dennett02,dennett03}. Such annual modulation has been observed
in a few IDV sources, as mentioned by \citet{gabanyi07}.

Very recently, the flat-spectrum radio quasar J~1128+5925 was found to show
strong IDV at centimeter wavelengths, and its IDV timescale displays an annual
modulation \citep{gabanyi07}. Therefore, its IDV may be caused by ISS. In
optical, there is no previous report on its variability. In order to know the
properties of its optical variability and to make a comparison to those of
its radio variability, we performed an optical monitoring program on this
object in 2007 May. Here we present the results.

\section{OBSERVATIONS AND DATA REDUCTION}
The monitoring was performed with a 60/90 cm Schmidt telescope at
Xinglong Station, National Astronomical Observatories of China. The Schmidt
telescope, is equipped with a $4096\times4096$
E2V CCD, which has a pixel size of $12\,\micron$ and a spatial resolution of
$1.3\,\arcsec\rm{pixel}^{-1}$. When the system is used for blazar monitoring,
only the central $512\times512$ pixels are read out as an image. The
monitoring was done in the $R$-band with exposure times ranging from 300 to
480 s, and covered the period from 2007 May 5 to 29. Because of the weather
condition and observations of other targets, there are actually 11 nights of
data on J~1128+5925.

\begin{figure}
\plotone{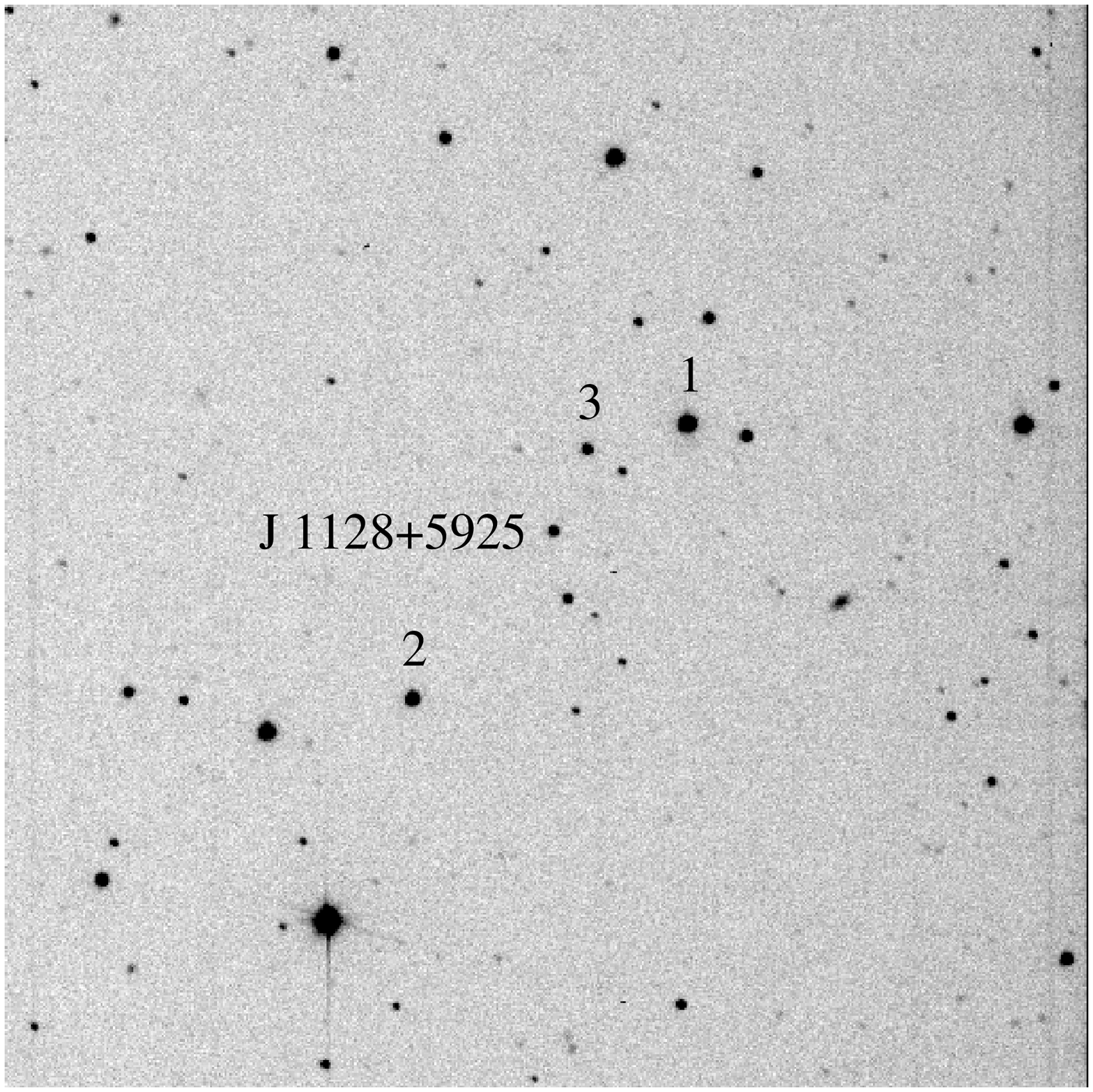}
\caption{Finding chart of J 1128+5925. The blazar and three comparison stars
are labeled.}
\label{F1}
\end{figure}

The data reduction procedures include positional calibration, bias subtraction,
flat-fielding, extraction of instrumental aperture magnitude, and flux
calibration. We adopted radii of the aperture and the sky annuli as 3, 7,
and 10 pixels, respectively, during the aperture photometry. Three stars
around J~1128+5925 were selected as comparison stars, as shown in the finding
chart in Figure~\ref{F1}. Here we used differential photometry. For each
frame, the instrumental magnitudes of the blazar and three comparison stars
were extracted. The brightness of the blazar was measured relative to the
average brightness of the three comparison stars. The differential magnitudes
of Star 3 (it has similar brightness to the blazar) relative to the average
of all three stars were also calculated to verify the stable fluxes of the
comparison stars and the accuracy of our measurements.

\section{RESULTS}
\begin{figure}
\plotone{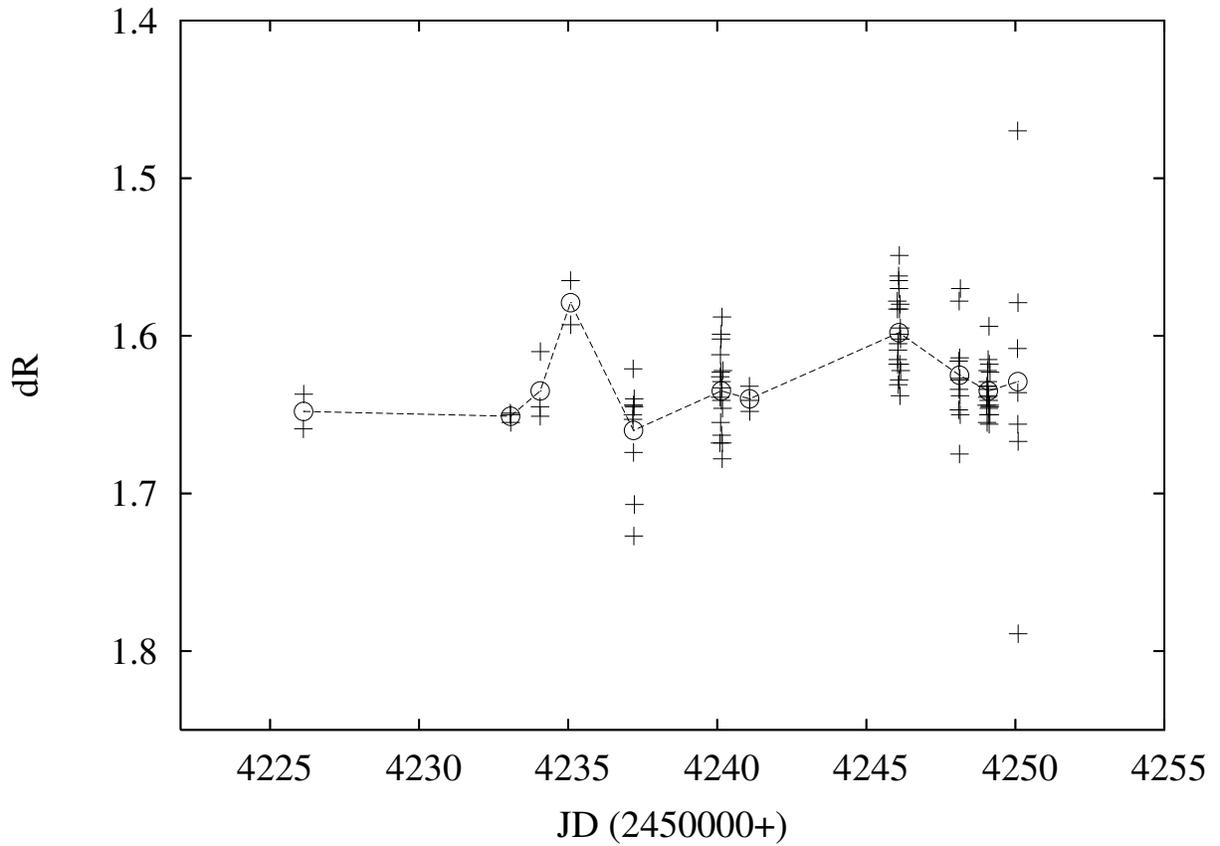}
\caption{The light curve of all nights. The plus symbols are individual
measurements, while the open circles and dashed line mark the nightly average
light curve. The brightest and faintest points on the last night may be
spurious measurements and were excluded from further analysis. The errors are
typical at 0.02 mags and are not plotted just for clarity.}
\label{F2}
\end{figure}

The light curve of the whole monitoring period is shown in Figure~\ref{F2}.
It is clear that there is no strong internight variation, except at
JD~2,454,235, where the strongest internight variation occurred with an
amplitude of $0.069\pm0.023$ mags when adopting the definition of \citet{wu07}.

\begin{figure}
\plottwo{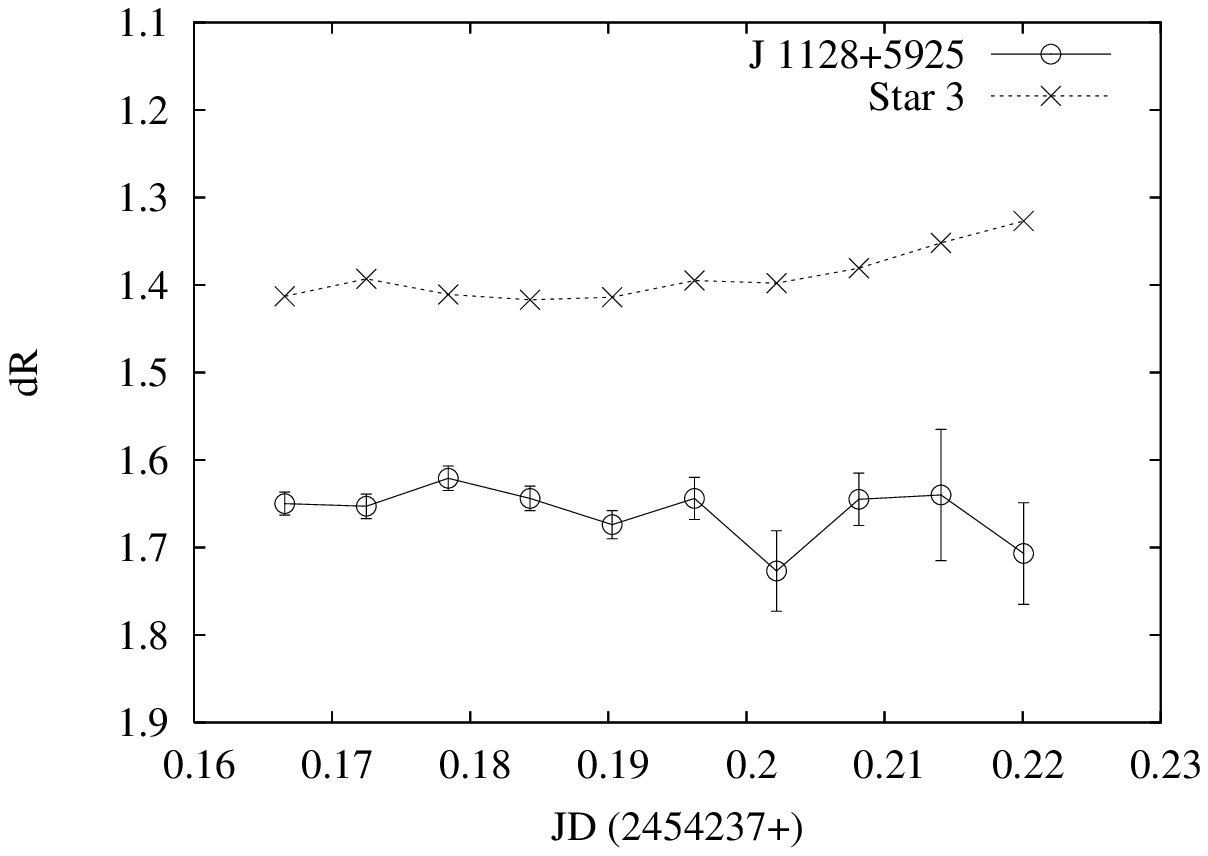}{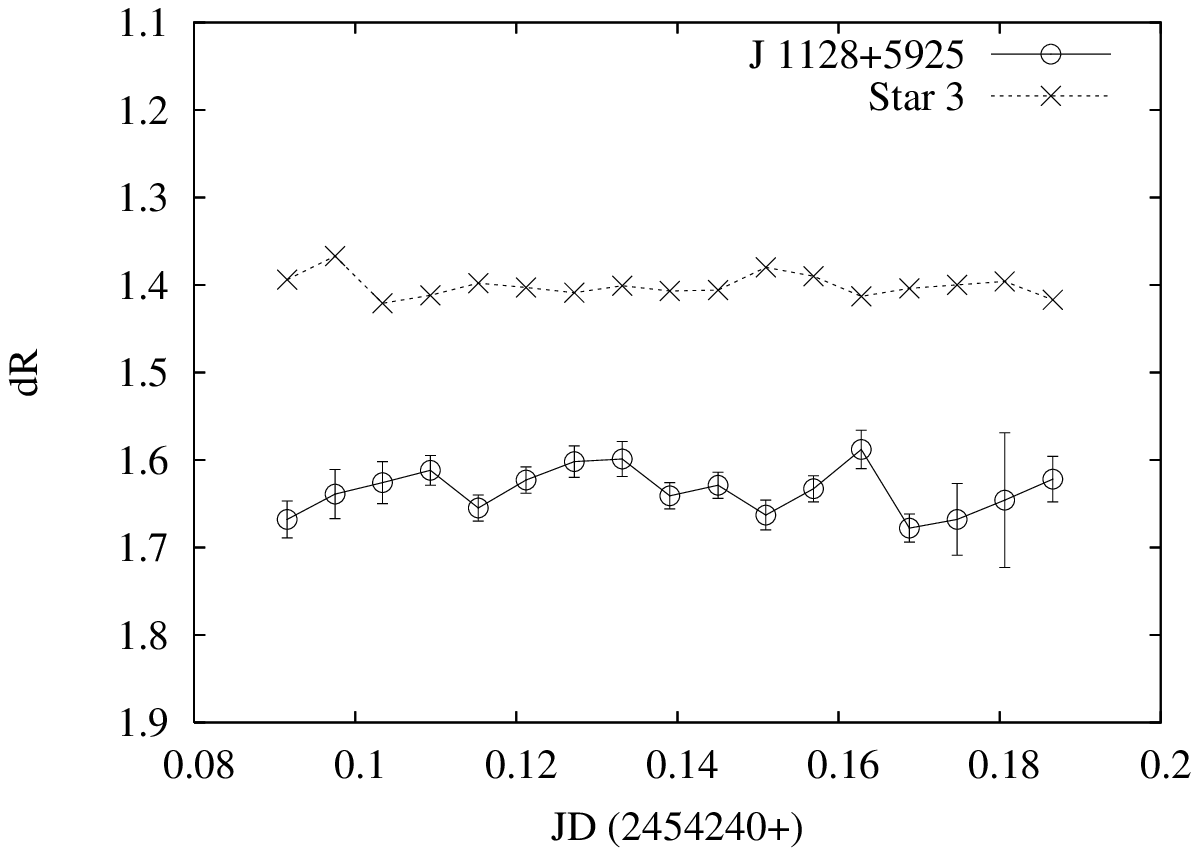}
\plottwo{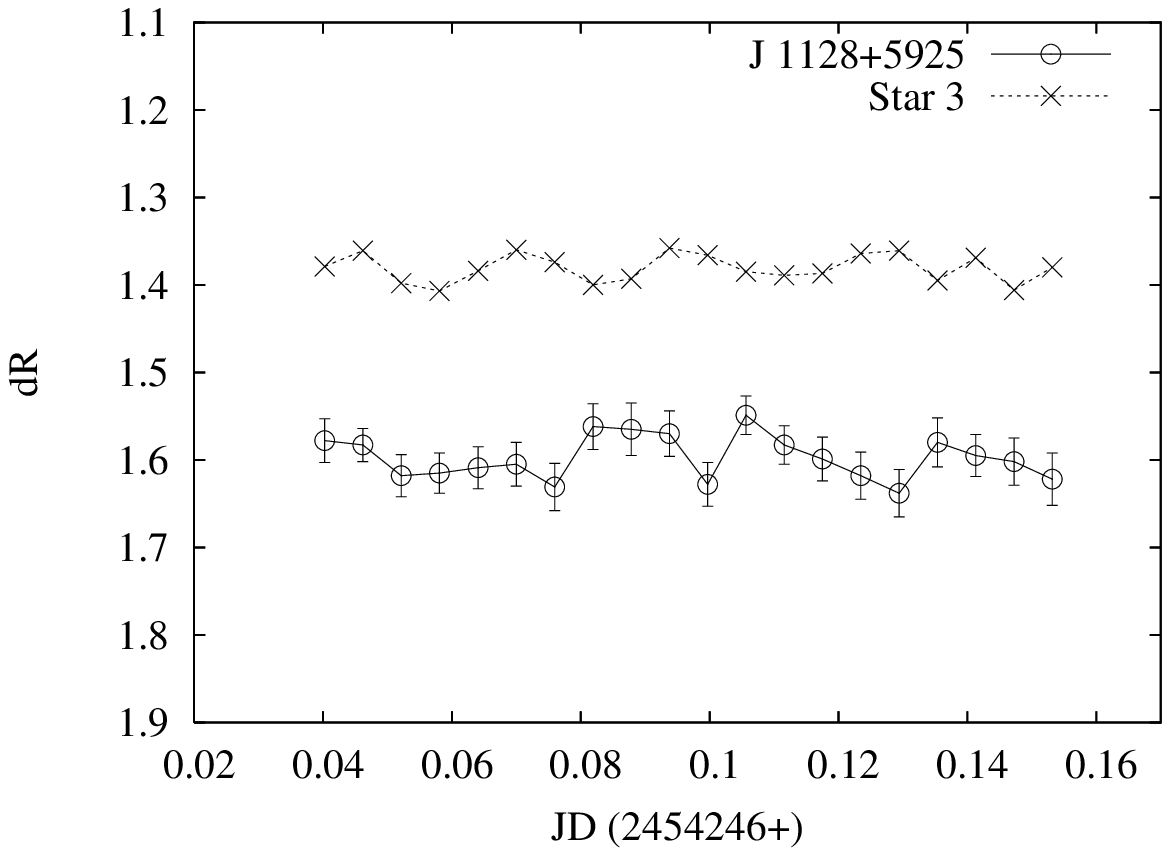}{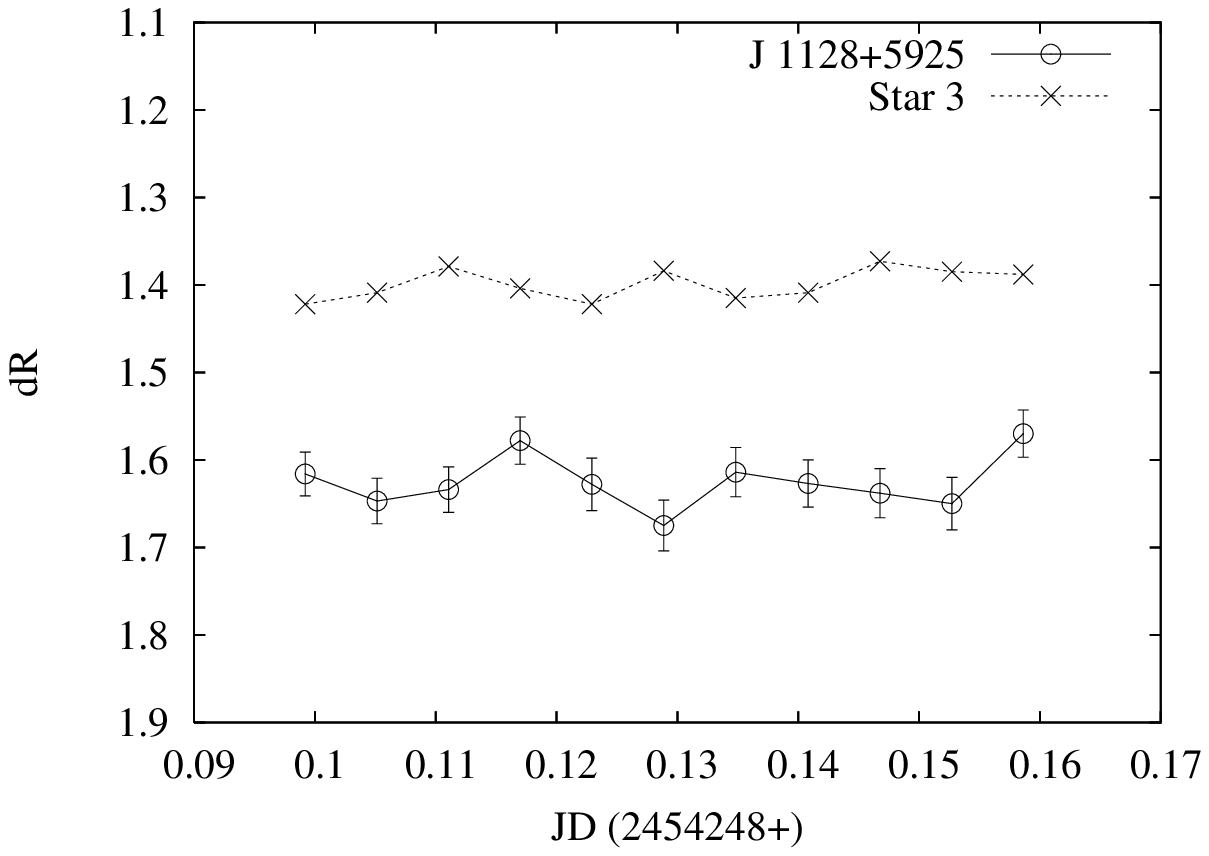}
\epsscale{0.45}
\plotone{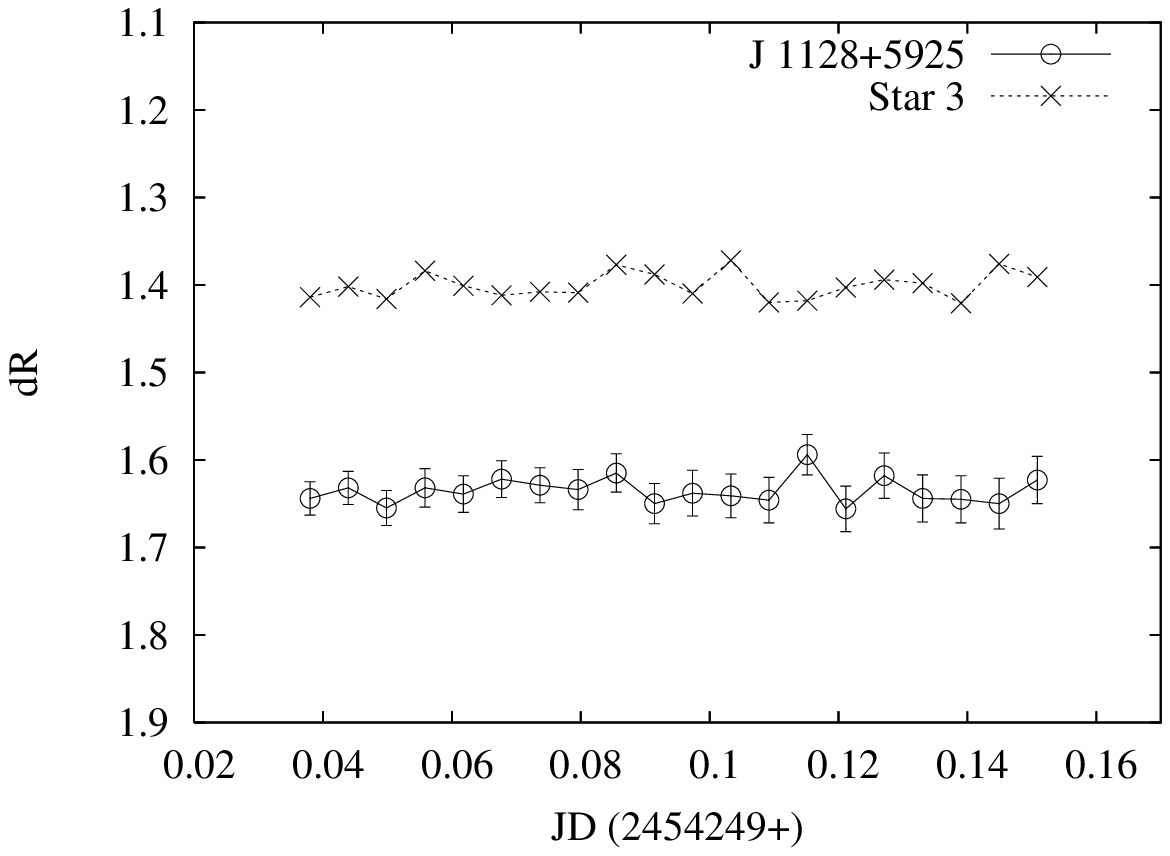}
\caption{Intranight light curves on JDs~2,454,237, 2,454,240, 2,454,246,
2,454,248, and 2,454,249, which are five most intensively monitored nights.
The open circles and solid lines show the light curves of the blazar. The
crosses and dotted lines show the light curves of Star 3.}
\label{F3}
\end{figure}

Figure~\ref{F3} displays the intranight light curves on five most intensively
monitored nights. The differential magnitudes of Star 3 are very stable at
around $dR\sim1.4$ with the exception of the last four points on JD~2,454,237.
On this night, the relatively low signal-to-noise ratio of the last four
CCD images resulted in the observed increasing brightness of Star 3, and
the large error bars on the light curve of the blazar.

In all five intranight light curves, the brightness of the object shows only
some small-amplitude oscillations around a constant average, and do not
demonstrate any clear tendency to become brighter or fainter. In fact, the
apparently random, small-amplitude oscillations on very short timescales
suggest that the oscillations may be due mainly to noise. A quantitative
assessment was performed on whether or not the object was variable on these
five nights. We employed the frequently used criteria adopted by, e.g.,
\citet{jang97}, \citet{stalin06}, and \citet{hu06}. We follow the convention
of defining C such that
$C=\sigma_{\rm B}/\sigma_{\rm S}$, where $\sigma_{\rm B}$ is the standard
deviation of the magnitudes of the blazar and $\sigma_{\rm S}$ is that of
the comparison star. When $C\geqslant2.576$, the object can be claimed to be
variable at the 99\% confidence level. Table~1 lists the results. All five $C$'s
are less than 2.0, implying that J~1128+5925 was not variable on these five
nights.

Although the monitoring periods on individual nights are only 1.28 to 2.71
hours long (see Table~1), the average observed brightness actually continued
on longer timescales, and in some cases may have extended to the next night
(e.g., from JD~2,454,240 to 41 and from JD~2,454,248 to 50). Therefore,
J~1128+5925 did not show strong variability on internight timescale, and even
did not vary on intranight timescale in this period of time. 

\section{CONCLUSIONS AND DISCUSSIONS}
We performed an optical monitoring program on J~1128+5925 in the $R$-band
from 2007 May 5 to 29. Our monitoring results indicate that in this period
J~1128+5925 only showed trivial optical variability on internight timescale,
and did not show any clear intranight variability. Either this object was in
a quiescent state in optical regimes in this period, or it is intrinsically
not as active at optical as it is at radio wavelengths.

Some blazars that show strong IDV in radio regimes also display rapid and
strong variability in optical regimes, such as S5~0716+714 \citep[e.g.,][]
{wu05,wu07,montagni06,pollock07}. This doesn't seem to be the case for
J~1128+5925. This object exhibits strong IDV at radio wavelengths, but not
at optical wavelengths. It is easy to explain this difference if the optical
and radio variabilities come from different origins: The optical variability
may be intrinsic to the source (the ISS cannot change the optical flux), while
the radio variability is mainly the result of ISS, as implied by the
observations of \citet{gabanyi07}.

We present the first report on the optical variability of J~1128+5925 in
this paper. However, because of the solar conjunction and observations of
other targets with the telescope, our monitoring did not last long. More
observations are needed to know whether or not this object is always
optically quiescent. Multi-band optical monitoring is also necessary to
constrain its optical variability in more detail. Of particular interest is
to carry out simultaneous optical and radio monitorings on this object in
order to make a more direct comparison between the variabilities at these
two wavelengths. Future campaigns can investigate whether there is
correlated optical IDV when strong radio IDV is observed. If such correlations
are detected, it would be strong evidence that both the optical and radio
variability structures are intrinsic to the source, as in the case of
S5~0716+714 \citep{quirren91,wagner95,wagner96}. The broadband variability
is also helpful to derive for this object some basic parameters, such as the
mass of the central supermassive black hole, the boosting factor of the
relativistic jet, etc \citep[e.g.,][]{fan05}.

\begin{acknowledgements} 
We thank the anonymous referee for insightful suggestions and comments that
helped to improve the presentation of this paper very much. This work has
been supported by the Chinese National Natural Science Foundation grants
10603006, 10473012, 10573020, 10633020.
\end{acknowledgements} 

%\clearpage

\clearpage

\clearpage

\begin{deluxetable}{ccccc}
\tablewidth{0pt}
\tablecaption{Statistics on Five Intranight Light Curves}
\tablehead{ \colhead{Julian Date} & \colhead{N} & \colhead{Duration (hrs)} &
\colhead{$C$} & \colhead{Var?} \\ }
\startdata
2,454,237 & 10 & 1.28 & 1.11 & N \\
2,454,240 & 17 & 2.28 & 1.96 & N \\
2,454,246 & 20 & 2.71 & 1.58 & N \\
2,454,248 & 11 & 1.43 & 1.72 & N \\
2,454,249 & 20 & 2.71 & 1.01 & N \\
\enddata
\tablecomments{Column 1 gives the Julian Date. Columns 2 and 3 list the
number and duration of observations on a night. Column 4 presents the $C$
value (see text), and column 5 indicates whether the object is variable (V)
or not (N).}
\end{deluxetable}


\begin{thebibliography}{}
\bibitem[Dennett-Thorpe \& de Bruyn(2002)]{dennett02}Dennett-Thorpe, J. \& de
Bruyn, A. G. 2002, \nat, 415, 57
\bibitem[Dennett-Thorpe \& de Bruyn(2003)]{dennett03}Dennett-Thorpe, J. \& de
Bruyn, A. G. 2003, \aap, 404, 113
\bibitem[Fan(2005)]{fan05}Fan, J. H. 2005, \aap, 436, 799
\bibitem[Gab\'anyi et al. (2007)]{gabanyi07}Gab\'anyi, K, \'E, et al. 2007,
\aap, 470, 83
\bibitem[Heeschen(1984)]{heeschen84}Heeschen, D. S. 1984, \aj, 89, 1111
\bibitem[Heeschen et al.(1987)]{heeschen87}Heeschen, D. S., Krichbaum, Th.,
Schalinski, C. J., \& Witzel, A. 1987, \aj, 94, 1493
305, 42
\bibitem[Hu et al.(2006)]{hu06}Hu, S. M., Wu, J., Zhao, G., \& Zhou, X.
2006, \mnras, 373, 209
\bibitem[Jang \& Miller(1997)]{jang97}Jang, M. \& Miller, H. R. 1997, \aj,
114, 565
\bibitem[Kellermann \& Pauliny-Toth(1969)]{keller69}Kellermann, K. I. \&
Pauliny-Toth, I. I. K. 1969, \apj, 155, L71
\bibitem[Montagni et al.(2006)]{montagni06}Montagni, F., Maselli, A., Massaro,
E., Nesci, R., Sclavi, S., \& Maesano, M. 2006, \aap, 451, 435
\bibitem[Pollock et al.(2007)]{pollock07}Pollock, J. T., Webb, J. R., \&
Azarnia, G. 2007, \aj, 133, 487
\bibitem[Quirrenbach et al.(1991)]{quirren91}Quirrenbach, A., Witzel, A.,
Wagner, S., et al. 1991, \apj, 372, L71
\bibitem[Stalin et al.(2006)]{stalin06}Stalin, C. S., Gopal-Krishna, Sagar,
R., Wiita, P. J., Mohan, V., \& Pandey, A. K. 2006, \mnras, 366, 1337
\bibitem[Wagner \& Witzel(1995)]{wagner95}Wagner, S. J. \& Witzel, A. 1995,
\araa, 33, 163
\bibitem[Wagner et al.(1996)]{wagner96}Wagner, S. J., et al. 1996, \aj, 111,
2187
\bibitem[Witzel et al.(1986)]{witzel86}Witzel, A., Heeschen, D. S., Schalinski,
C., \& Krichbaum, Th. 1986, Mitteilungen der Astronomischen Gesellschaft, 65,
239
\bibitem[Wu et al.(2005)]{wu05}Wu, J., Peng, B., Zhou, X., Ma, J., Jiang, Z.,
\& Chen, J. 2005, \aj, 129, 1818
\bibitem[Wu et al.(2007)]{wu07}Wu, J., Zhou, X., Ma, J., Wu, Z., Jiang, Z.,
\& Chen, J. 2007, \aj, 133, 1599
\end{thebibliography}
\end{document}